 \documentclass[twocolumn,aps,showpacs,pre,amsmath]{revtex4}

\usepackage{amssymb}
\usepackage{graphicx}
\usepackage{dsfont}

\begin{document}

\title{Strain-induced conduction gap in vertical devices \\ made of twisted graphene layers}

\author{V. Hung Nguyen$^{1,2}$\footnote{E-mail: hung@iop.vast.ac.vn}, Huy-Viet Nguyen$^2$, J. Saint-Martin$^1$, and P. Dollfus$^1$} \address{$^1$Institut d'Electronique Fondamentale, UMR8622, CNRS, Universit$\acute{e}$ Paris Sud, 91405 Orsay, France \\ $^2$Center for Computational Physics, Institute of Physics, Vietnam Academy of Science and Technology, P.O. Box 429 Bo Ho, 10000 Hanoi, Vietnam}

\begin{abstract}
 We investigate the effects of uniaxial strain on the transport properties of vertical devices made of two twisted graphene layers, which partially overlap each other. We find that because of the different orientations of the two graphene lattices, their Dirac points can be displaced and separated in the $k-$space by the effects of strain. Hence, a finite conduction gap as large as a few hundred \emph{meV} can be obtained in the device with a small strain of only a few percent. The dependence of this conduction gap on the strain strength, strain direction, transport direction and twist angle are clarified and presented. On this basis, the strong modulation of conductance and significant improvement of Seebeck coefficient are shown. The suggested devices therefore may be very promising for improving applications of graphene, e.g., as transistors or strain and thermal sensors.
\end{abstract}

\pacs{xx.xx.xx, yy.yy.yy, zz.zz.zz}
\maketitle

The continuing interest in graphene, a two-dimensional (2D) monolayer arrangement of carbon atoms, as conducting material is one of the most striking trends of research in solid-state and applied physics over the last decade \cite{neto09,ywu013,yazy10,bala11,shar13,novo12,schw10}. It is due in particular to the specific band structure of this material, i.e. with gapless conical shape at the six edge corners of the hexagonal Brillouin zone and the Dirac character of low-energy excitations, which leads to many peculiar effects as relativistic-like behavior of charge carriers, finite value of the conductivity at zero density, unusual quantum Hall effect, etc \cite{neto09}. It is due also to outstanding properties as high carrier mobility, small spin-orbit coupling, high thermal conductivity and excellent mechanical properties, which make it very promising for a broad range of applications.
However, in the view of operation of electronic devices, graphene still has serious drawbacks associated with the lack of an energy bandgap in its electronic structure. In particular, graphene transistors have low ON/OFF ratio and poor current saturation \cite{meri08}. Many efforts of bandgap engineering in graphene have been made to solve these issues. For instance, the techniques as cutting 2D graphene sheet into narrow nanoribbons \cite{yhan07}, depositing graphene on hexagonal boron nitride (hBN) substrate \cite{khar11}, nitrogen-doped graphene \cite{lher13}, applying an electric field perpendicularly to Bernal-stacking bilayer graphene \cite{zhan09}, graphene nanomeshes \cite{jbai10}, using hybrid graphene/hBN \cite{fior12} or vertical graphene channels \cite{brit12} have been explored. Though being certainly promising options, each of these technique still has its own issues. Hence, the bandgap engineering is still a timely and desirable topic at the moment for the development of graphene in nanoelectronics.

Besides these points mentioned above, graphene is also an attractive material for flexible electronics since it is able to sustain a much larger (i.e., $> 20\%$ \cite{shar13}) strain than other semiconductors. Recently, some techniques \cite{garz14,shio14} to generate extreme strain in graphene in a controlled and nondestructive way have been also explored. Interestingly, strain engineering has been suggested to be an approach to modulate efficiently the electronic structure of graphene nanomaterials. On this basis, many interesting electrical, optical and magnetic properties induced by strain have been investigated, e.g., see refs. \cite{pere09a,pere09b,cocc10,yalu10,kuma12,baha13,hung14a,chun14,pere10,gxni14,guin10,tlow10,zhai11}. Remarkably, although the slightly strained (i.e., a few percent) 2D graphene remains semimetallic \cite{pere09b}, strain has been demonstrated as a technique for strongly improving the applications of some particular graphene channels, for instance, graphene nanoribbons with a local strain \cite{yalu10,baha13}, graphene with grain boundaries \cite{kuma12} and graphene strain junctions \cite{hung14a,chun14}.

Recently, the interest of the graphene community has been also oriented toward the investigation of specific graphene systems, named twisted few-layer graphene lattices. They are actually few-layer graphene lattices where one layer is rotated relative to another layer by an arbitrary angle and can form a Moir\'{e} pattern. These graphene lattices often appear in the thermal decomposition of the C-face of SiC or in the copper-assisted growth using the chemical vapor deposition method, e.g., see refs. \cite{heer07,hass08,caro11,luic11,have12,cclu13}. In the twisted graphene bilayer, the bandstructure changes dramatically \cite{sant07,lais10,guli10}, compared to that of monolayer or Bernal/AA stacking bilayer systems. In addition to the existence of linear dispersion in the vicinity of K-points, saddle points emerge at the crossing of Dirac cones, yielding van Hove singularities in the density of states at low energies, and the Fermi velocity to be remarkably renormalized. Moreover, the phonon transport in these systems also exhibits a strong dependence on the twist angle \cite{coce13}. Motivated by recent studies \cite{kuma12,hung14a,chun14} of the strain effects to generate/modulate the conduction gap in graphene channels, we investigate in this work the effects of uniaxial strain on the charge transport in vertical devices made of two twisted graphene layers as schematized in Fig. 1. The idea is as follows. In any hetero-channels made of different graphene sections where their Dirac cones are located at different positions in the $k-$space, the transmission probability (and hence conductance/current) shows finite energy gaps, i.e., conduction gaps, even though these graphene sections are still metallic (similarly, see the detailed explanation in \cite{kuma12,chun14}). This feature is also expected to be observed here because the graphene lattices in the left and right sides have different orientations and hence their electronic structures should be, in principle, different when a strain is applied. Moreover, compared to the strain heterochannels \cite{hung14a,chun14} and vertical devices \cite{brit12} previously studied, the advantages of these devices come from the use of a uniform strain and graphene materials only, which can make it a quite simple option for the fabrication process.

\begin{figure}[!t]
\centering
 \includegraphics[width=3.3in]{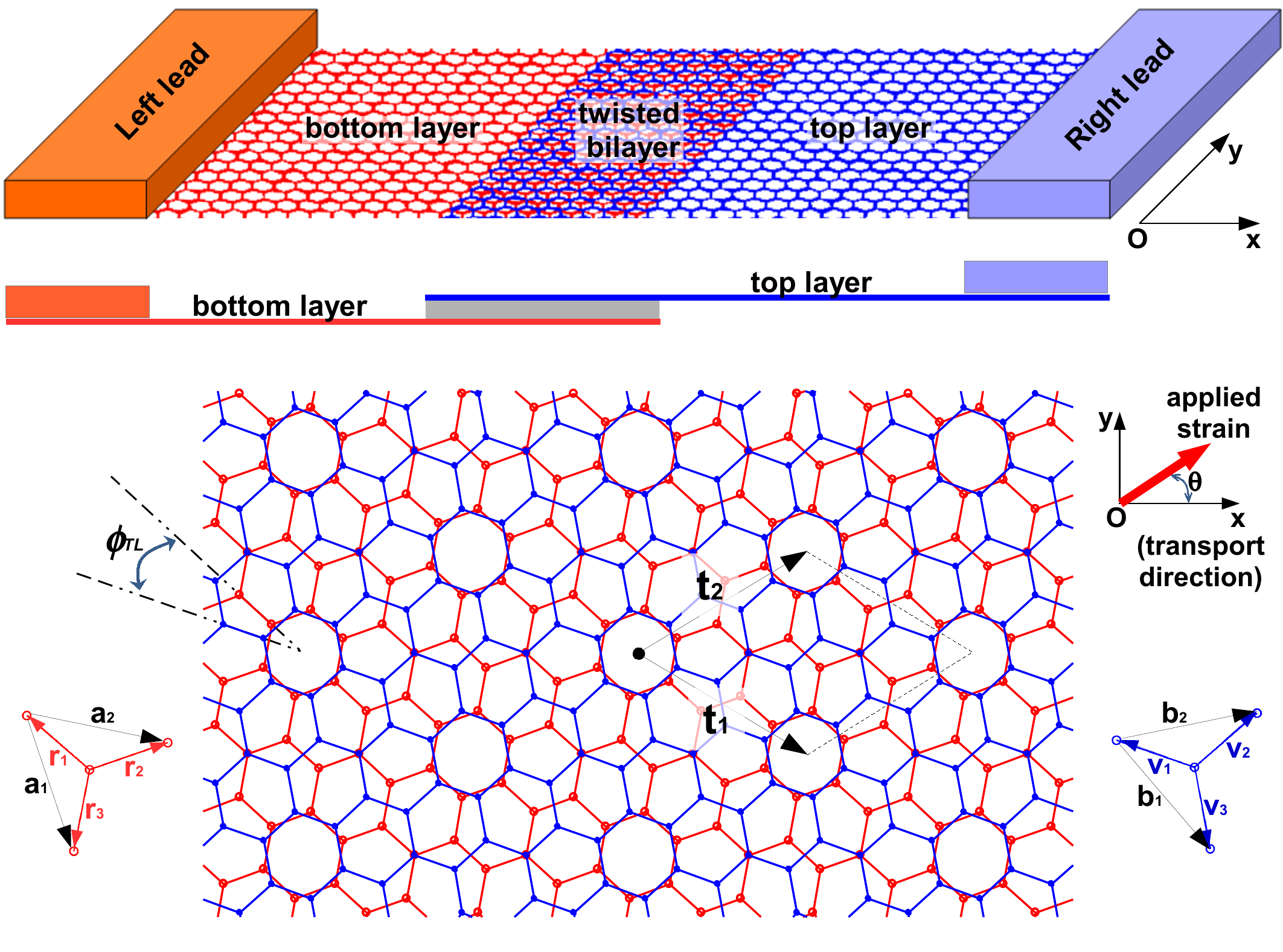}
\caption{Schematic of vertical graphene devices investigated in this work (top) and their side view (middle). The bottom shows the top view of a typical twisted graphene bilayer lattice.}
\label{fig_sim0}
\end{figure}
For the investigation of charge transport in the proposed devices, we employed atomistic tight-binding calculations as in \cite{pere09b,lais10,sant12,hung14a,chun14}. Here, we assume that: (i) two (bottom and top) graphene sheets partially overlap each other and the transport (i.e., Ox) direction is perpendicular to this overlap section as shown in Fig. 1; (ii) the top sheet is rotated relative to the bottom one by a commensurate angle $\phi_{TL}$; (iii) a uniformly uniaxial strain is applied in the in-plane direction with an arbitrary angle $\theta$ with respect to the transport direction. The commensurate angles are determined by $\cos {\phi _{TL}} = ({n^2}/2 + 3mn + 3{m^2})/({n^2} + 3mn + 3{m^2})$ \cite{sant12}, where \emph{n} and \emph{m} are coprime positive integers. The primitive vectors shown in Fig. 1 are determined as follows: ${\vec t_1} = m{\vec a_1} + \left( {n + m} \right){\vec a_2}$ and ${\vec t_2} =  - \left( {n + m} \right){\vec a_1} + \left( {n + 2m} \right){\vec a_2}$ if gcd(\emph{n},3) = 1; ${\vec t_1} = \left( {\frac{n}{3} + m} \right){\vec a_1} + \frac{n}{3}{\vec a_2}$ and ${\vec t_2} =  - \frac{n}{3}{\vec a_1} + \left( {\frac{{2n}}{3} + m} \right){\vec a_2}$ if gcd(\emph{n},3) = 3 [where gcd(\emph{p,q}) is the greatest common divisor of \emph{p} and \emph{q}]. For simplicity, throughout the work, unless otherwise stated the transport direction is chosen to be parallel to the vector ${\vec L_0} = {\vec t_1} + {\vec t_2}$. The strain causes changes in the $C-C$ bond vector $\vec r_{ij}$ according to ${\vec r_{ij}}\left( \sigma\right) = \left\{ {\mathds{1} + {M_s}\left( {\sigma ,\theta } \right)} \right\}{\vec r_{ij}}\left( 0 \right)$ with the strain tensor
\begin{equation}
  {M_s}\left( {\sigma ,\theta } \right) = \sigma \left[ {\begin{array}{*{20}{c}}
{{{\cos }^2}\theta  - \gamma {{\sin }^2}\theta }&{\left( {1 + \gamma } \right)\sin \theta \cos \theta }\\
{\left( {1 + \gamma } \right)\sin \theta \cos \theta }&{{{\sin }^2}\theta  - \gamma {{\cos }^2}\theta }
\end{array}} \right] \nonumber
\end{equation}
where $\sigma$ represents the strain and $\gamma \simeq 0.165$ is the Poisson ratio \cite{blak70}. Taking into account the strain effects, the hopping parameters of tight binding Hamiltonian are adjusted accordingly as in \cite{pere09b}. To compute the transport quantities (transmission probability and conductance) and extract the value of conduction gap, we used the non-equilibrium Green's function technique and the bandstructure analysis, as described in \cite{hung14a,chun14}. The conduction gap mentioned here is essentially the energy window within which the Fermi energy can be varied (e.g., by applying and tuning a back gate voltage) and the channel remains insulating. Note that due to the lattice symmetry, the twisted graphene bilayer lattices with $\phi_{TL} = 30^\circ - \alpha$ and $30^\circ + \alpha$ are equivalent while the applied strain of angle $\theta$ is identical to that of $\theta + 180^\circ$. It hence limits our investigation to $\phi_{TL} \in [0,30^\circ]$ and $\theta \in [-90^\circ,90^\circ]$.

\begin{figure*}[!t]
\centering
\includegraphics[width=5.6in]{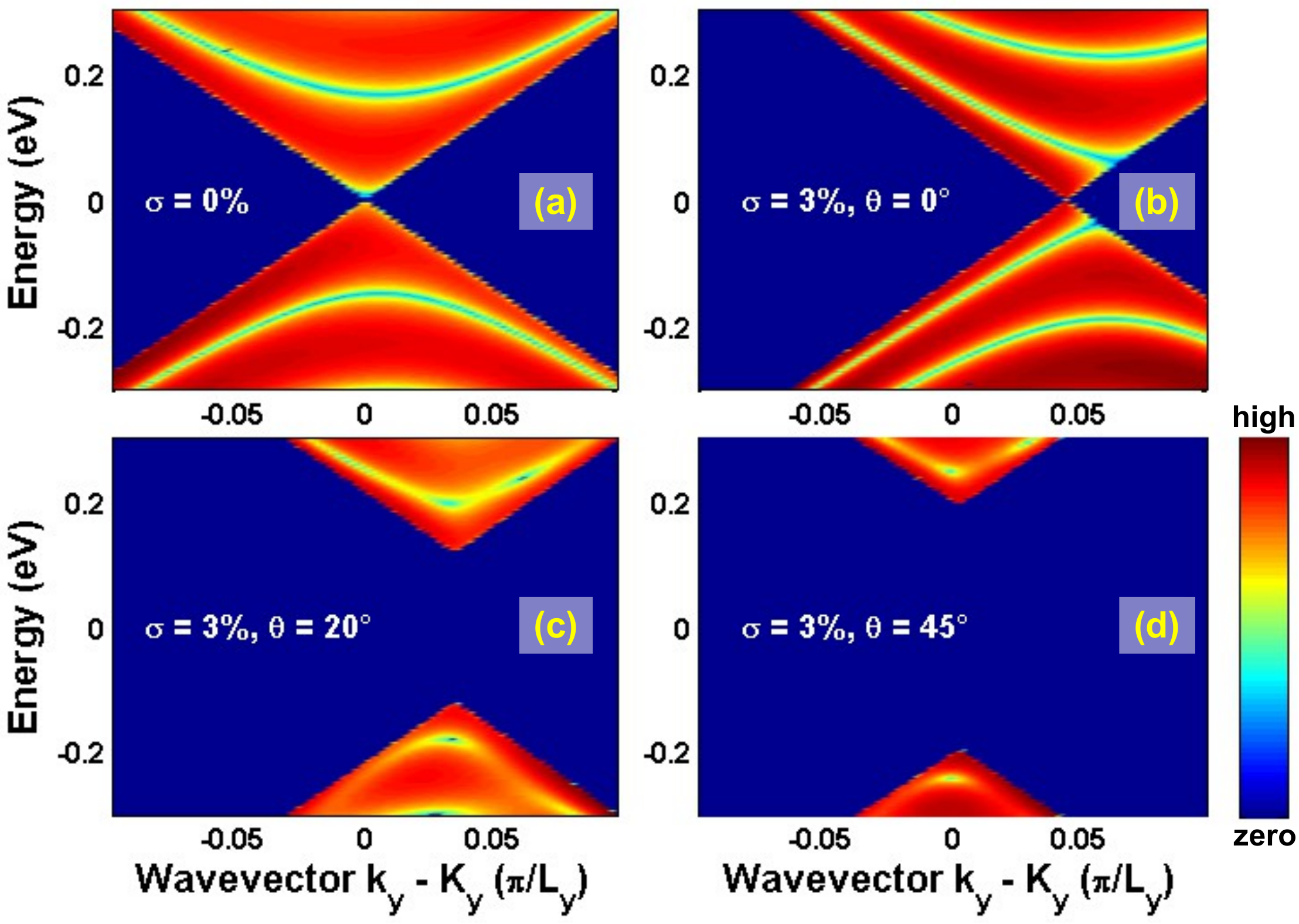}
\caption{($E-k_y$) maps of transmission coefficient of considered devices around the Dirac point with different applied strains. The twist angle $\phi_{TL} = 21.8^\circ$ is considered and $L_y \equiv \|{\vec t_2} - {\vec t_1}\|$ is the size of unit cells in the Oy-direction.}
\label{fig_sim1}
\end{figure*}
In Fig. 2, we present $E-k_y$ maps showing the main effects of strain on the transmission coefficient of considered devices in the case of $\phi_{TL} \simeq 21.8^\circ$ (i.e., $n = m = 1$). First, the device remains metallic with a zero conduction gap in the case of unstrained layers (see Fig. 2(a)). This is because the Dirac cones of graphene sections in the left and right sides are still located at the same $k_y-$position, i.e., at the K-point. The strain can induce a displacement of Dirac cones from the K-point \cite{pere09b,chun14} and a finite gap can open in the device if the Dirac cones of two such graphene sections are separated along the $k_y$-direction, similarly to what was explained in \cite{chun14}. Therefore, the transport picture is dramatically changed as shown in Figs. 2(b,c,d). In Fig. 2(b), although the Dirac cones are displaced, the device is still metallic with a zero conduction gap. This is essentially explained by the fact that the system is symmetric with respect to the overlap region (i.e., the Oy direction) even the strain of 3$\%$ with $\theta = 0^\circ$ is applied. Because of this symmetry, the Dirac cones of left and right sections are still located at the same $k-$point, which explains the zero gap observed. This symmetry can be broken when the direction of applied strain changes, leading to the opening of a finite conduction gap (see Figs. 2(c,d)). Indeed, finite gaps of $\sim$ 240 meV and 390 meV are achieved for the strain angles $\theta = 20^\circ$ and $45^\circ$, respectively. Thus, these data show two important features: (i) the strain can induce a finite conduction gap in the device under study and (ii) besides the strain strength, the gap is strongly dependent on the strain direction. Similar features have been also reported in \cite{chun14} for monolayer graphene strain junctions.

In Fig. 3, we display a picture showing the properties of conduction gap in the device discussed above with respect to the strain strength, strain direction and also the transport direction. The data in Fig. 3(a) was obtained for the case where the transport direction is parallel to $\vec L_0$ and the strain strength $\sigma = 3\%$ is applied. It is shown that the conduction gap is a function of strain direction $\theta$ with two peaks at $\theta \simeq \pm 45^\circ$ and zero values for $\theta = 0^\circ$ and $\pm 90^\circ$. The reason why the gap is zero at $\theta = \pm 90^\circ$ is essentially similar to that for which the zero gap is observed at $\theta = 0^\circ$ explained above. Figs. 3(b,c) presents the maps of conduction gap with respect to the strain strength and its applied direction in the tensile and compressive cases, respectively. We find that (i) the gap almost linearly increases with the strain strength; (ii) for a given strength, the compressive strain gives a larger gap than the tensile one; (iii) differently from the strain junctions in ref. \cite{chun14}, both kinds of strain gives a similar dependence of conduction gap on the strain direction. Finally, since it is due to the separation of Dirac cones in the $k_y$-direction, the conduction gap is also strongly dependent on the direction of carrier transport. In Fig. 3(a), we additionally display the data obtained when the transport direction is parallel to $\vec L_1$ ($\equiv 4\vec t_1 - \vec t_2$) and $\vec L_2$ ($\equiv 5\vec t_1 + \vec t_2$), compared to the case of $\vec L_0$. Note that the vectors $\vec L_1$ and $\vec L_2$ actually correspond to the armchair direction of top and bottom layers, respectively. Our calculations show that in general, a finite conduction gap can always be observed but its dependence on the strain direction is dramatically changed when changing the transport direction. Indeed, as seen in Fig. 3(a), the $E_{gap} \left( \theta  \right)$ function exhibits two similar peaks and two valleys in all cases. However, the position of these peaks and valleys changes when changing the direction of transport.
\begin{figure*}[!t]
\centering
\includegraphics[width=5.8in]{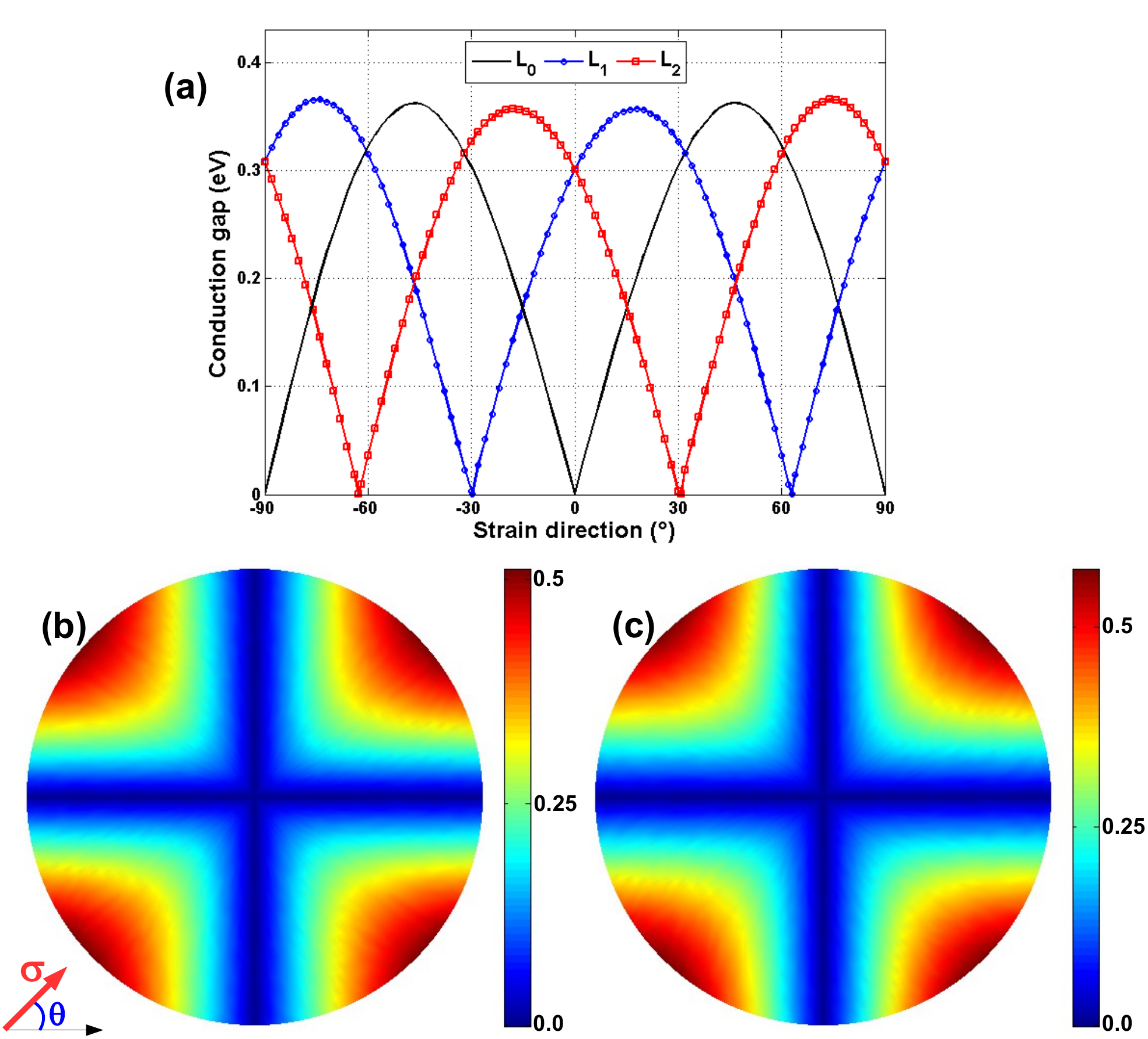}
\caption{(a) conduction gap as a function of strain direction ($\sigma = 3 \%$) with different transport directions, which are parallel to the vectors $\vec L_n$ (\emph{n} = 0,1,2) with $\vec L_0 = \vec t_1 + \vec t_2$, $\vec L_1 = 4\vec t_1 + \vec t_2$, and $\vec L_2 = 5\vec t_1 - \vec t_2$. The bottom shows the maps of conduction gap in the first case with respect to the ((b) tensile and (c) compressive) strain and its applied direction. In these maps, the radius from the central point represents the strain strength ranging from 0 (center) to 4 $\%$ (edge). The twisted layers with $\phi_{TL} = 21.8^\circ$ are considered here.}
\label{fig_sim2}
\end{figure*}

Next, we go to explore the properties of conduction gap with respect to the twist angle $\phi_{TL}$. In Fig. 4, we display the data obtained for two types of commensurate lattices \cite{sant12} corresponding to gcd(\emph{n},3) = 1 (lattice 01) and gcd(\emph{n},3) = 3 (lattice 02). In addition, we consider separately the two regimes of large $\phi_{TL}$ ($> 7.3^\circ$) in Fig. 4(a) and small $\phi_{TL}$ in Fig. 4(c). In the regime of large $\phi_{TL}$, we find that the similar $E_{gap} \left( \theta \right)$ behavior with finite peaks is observed for all cases investigated: $E_{gap}-$peaks are at $\theta \simeq \pm 45^\circ$ and zero values at $\theta = 0^\circ$ or $\pm 90^\circ$. More interestingly, two types of twisted lattices show opposite trends of conduction gap when increasing the twist angle $\phi_{TL}$: $E_{gap}-$peaks increase for the lattice type 02 while they are generally reduced in the case of the type 01. This phenomenon can be explained by the difference of the symmetry in these two lattice types. For illustration, we present a diagram in Fig. 4(b) showing the displacement and separation of Dirac cones of two graphene layers under strain of angle $\theta = 45^\circ$. The diagram shows that although their displacement are similar for all cases, the separation of Dirac cones of two graphene layers have different behaviors, especially along the $k_y$ direction. For the lattice type 02, this separation tends to increase when increasing the twist angle while it reduces for the lattice type 01. These properties basically explain the results obtained.

\begin{figure*}[!t]
\centering
\includegraphics[width=6.0in]{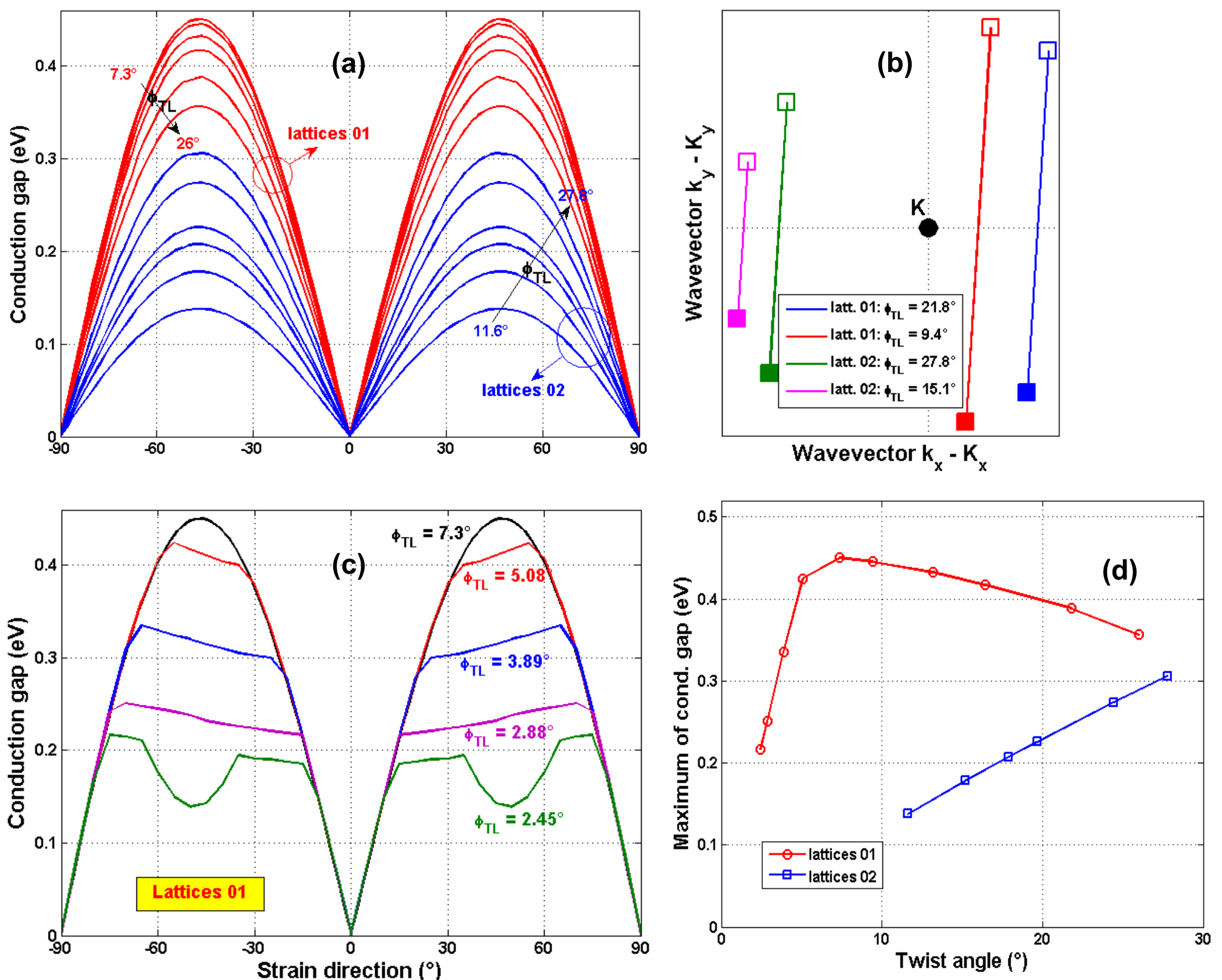}
\caption{Conduction gap as a function of strain direction different twist angles $\phi_{TL}$: large $\phi_{TL}$ (a) and small $\phi_{TL}$ (c). (b) diagram illustrating the strain-induced displacement of Dirac cones away from the \emph{K}-point in the case of $\theta = 45^\circ$. Open (filled) squares denote the Dirac cones of the bottom (top) graphene layers. (d) evolution of maximum values of conduction gap as a function of $\phi_{TL}$. Two lattice types 01 and 02 (see in the text) are considered and, everywhere, the strain $\sigma = 3 \%$ is applied.}
\label{fig_sim3}
\end{figure*}
In the regime of small $\phi_{TL}$, we find as shown in Fig. 4(c) another trend of $E_{gap} \left( \theta \right)$ in the case of lattices 01, i.e., $E_{gap}$ quite surprisingly reduces around $\theta = \pm 45^\circ$ when decreasing $\phi_{TL}$, compared to the data in Fig. 4(a). This feature can be explained as follows. On the one hand, the strain tends to enlarge the separation of Dirac cones of the two layers in the $k_y$-axis and hence $E_{gap}$ increases when tuning $\theta$ from $-90^\circ$ to $-45^\circ$ . On the other hand, the first Brillouin zone is small since the periodic cells are large in the lattices of small $\phi_{TL}$. As a consequence, the separation of Dirac cones is limited and its behavior can change (i.e., from increasing to decreasing or vice versa) when the Dirac cone of any layer reaches the edge of Brillouin zone. These properties essentially explain the sudden change of $E_{gap} \left( \theta \right)$ shown in Fig. 4(c). In the situation where the Dirac cones of only one layer can reach the edge of Brillouin zone when tuning $\theta$, we obtain $E_{gap} \left( \theta \right)$ as shown for $\phi_{TL} = 5.08^\circ$, $3.89^\circ$ and $2.88^\circ$. For smaller $\phi_{TL}$, the Dirac cones of both layers can sequentially reach the edge of Brillouin zone, we additionally observe the valleys of $E_{gap}$ around $\theta = \pm 45^\circ$, i.e., see the data for $\phi_{TL} = 2.45^\circ$ in Fig. 4(c). Note that, in the case of lattices 02, because both the maximum separation of Dirac cones induced by strain and the size of Brillouin zone tend to reduce when decreasing $\phi_{TL}$, the feature discussed here does not occur. The evolution of maximum values of $E_{gap}$ when changing $\phi_{TL}$ is summarized in Fig. 4(d). When decreasing $\phi_{TL}$, the maximum value of $E_{gap}$ decreases monotonically for lattices 02 while it has a peak but also tends to zero for lattices 01. On this basis, it is suggested that to safely achieve a finite conduction gap without requiring the good control of twist angle, designing devices with $\phi_{TL}$ around/close to $30^\circ$ should be a good option.

Now, we would like to discuss some possible applications for this type of heterochannels. First, the devices can be used to improve the performance of graphene transistors with the advantage of utilizing a uniform strain and graphene materials only, compared to the strain hetero-channels \cite{hung14a} and vertical devices \cite{brit12} previously studied. Indeed, as shown in Fig. 5(a), with a significant conduction gap, these devices can exhibit a very high ON/OFF current ratio, i.e., up to a few ten thousands for a small strain of only $\sim 3-4 \%$. Second, based on the strong sensitivity of conduction gap to a small strain and its applied direction, this type of devices could be an elementary component of strain sensors. Third, the opening of a finite conduction gap also provides the possibility of enhancing the thermoelectric properties of these devices by strain. As seen in Fig. 5(b), the Seebeck coefficient can reach high values of 600 to 800 \emph{$\mu$V/K} for a strain of $\sim 3-4 \%$ while it is only about a few tens of \emph{$\mu$V/K} in the unstrained case or in the pristine graphene \cite{zuev09}. This improvement may be significant for applications as thermal sensors \cite{herw86}. Additionally, although our calculations show that the power factor is still limited and hence the thermoelectric figure of merit $ZT$ is small, the phonon conductance is low in this type of devices and one can expect that $ZT$ can be further improved when including additional energy-gap engineering techniques, e.g., by creating a nanohole lattice in graphene layers as suggested in \cite{hung14b}.

\begin{figure}[!t]
\centering
 \includegraphics[width=3.3in]{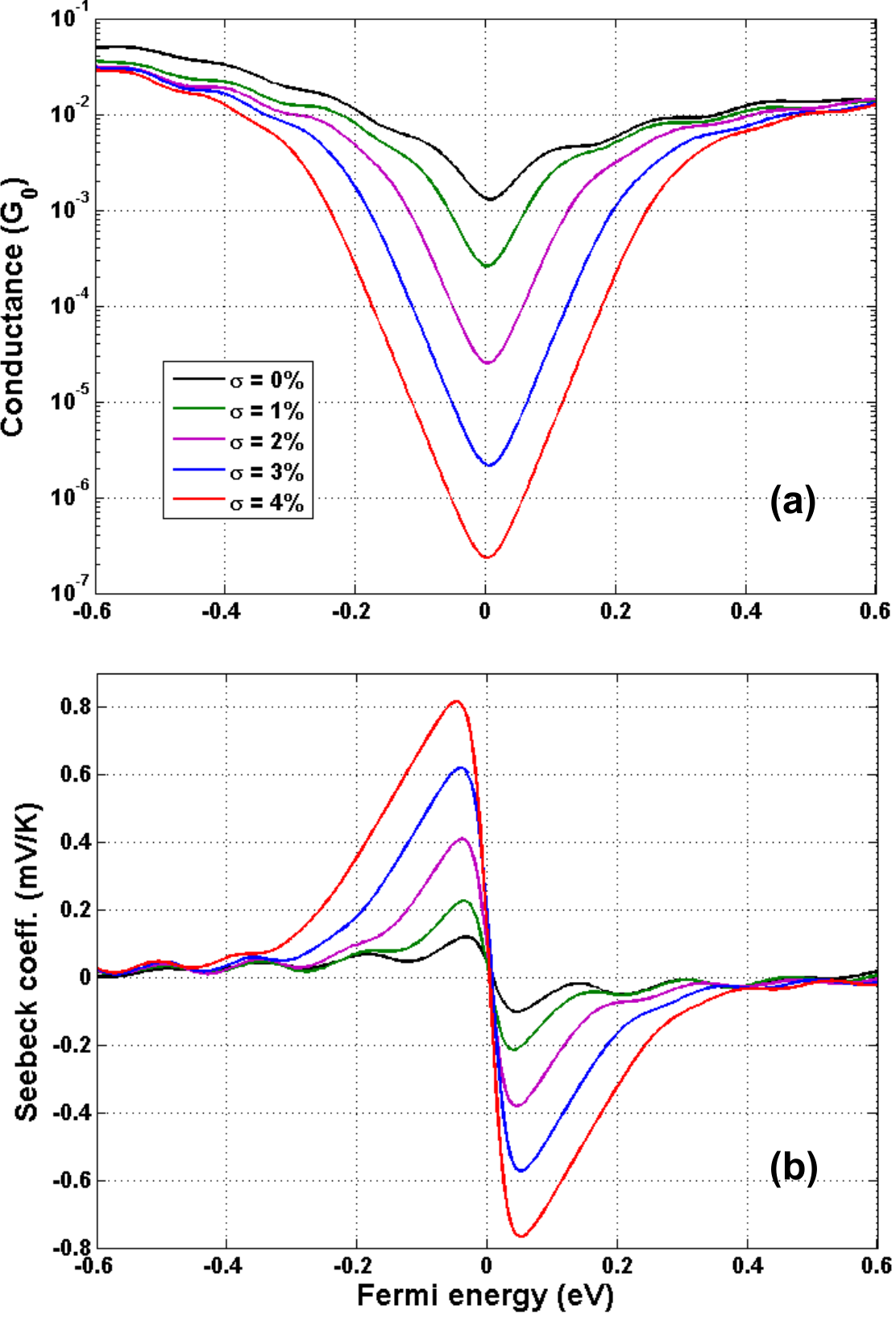}
\caption{(a) conductance and (b) Seebeck coefficient with different strain strengths at room temperature as a function of Fermi energy. The devices as in Fig. 2 and the strain direction $\theta = 45^\circ$ are considered. $G_0 = e^2W/hL_y$ with the channel width \emph{W} in the Oy direction.}
\label{fig_sim4}
\end{figure}
Finally, we would like to make some additional remarks. First, in this kind of devices, the overlap region between the two layers can have important effects on the transport properties. As in the study on vertical structures with Bernal stacking \cite{hung14b}, the size of this region determines, on the one hand, the coupling strength between layers and, on the other hand, the confinement effects (i.e., see in Fig. 2) because the electronic structure in the bilayer region is very different from that of left and right monolayer graphene sections. However, our calculations show that the change in the size of this overlap region does not dramatically change the ON-current (i.e., beyond the gap) except it can give rise to peaks and shallow valleys in the conductance as seen in Fig. 5. Second, we would like to notice that besides the use of a uniform strain, the vertical devices studied here have the additional advantage of being able to achieve the same values of conduction gap as the unstrained/strained graphene junctions in \cite{chun14}, but with smaller strain. For instance, a strain of $\sim$ 4$\%$ is enough to achieve a conduction gap of $\gtrsim$ 500 \emph{meV} while a strain of $\gtrsim$ 6$\div$7$\%$ is required in the latter channels. This improvement comes from the fact that the Dirac cones are displaced by strain in both left and right graphene sections, while in strained/unstrained junctions, the displacement of Dirac cones occurs only in the strained graphene section. Similar improvement can be achieved in the channels made of different strained graphene lattices, e.g., compressive/tensile strained junctions \cite{chun14}. However, the control of this complicated strain profile may be a practical issue. Third, besides the case of uniform strains studied here, similar effects can still be obtained in these devices if the strain is applied to only one layer or two different strains to two layers \cite{choi10}. In such cases, the properties of $E_{gap}$ should be, of course, strongly dependent on the strain configurations.

In conclusion, we have investigated effects of uniaxial strain on the transport properties of vertical devices made of two twisted graphene layers. It was shown that strain can induce the displacement of Dirac cones of both layers and because of their different orientations, these Dirac cones can be separated in the $k$-space. As a consequence, the device channel can be tuned from metallic to semiconducting by strain. A conduction gap larger than 500 \emph{meV} can be achieved in the device with a small strain of only $\sim$ 4$\%$. The dependence of this conduction gap on the strain strength, strain direction, transport direction and twist angle has been clarified. The twist angle $\phi_{TL} \simeq 30^\circ$ is the good option for a large conduction gap, which is less sensitive to the different types of twisted layers. On this basis, an ON/OFF current ratio as high as a few ten thousands and a strong improvement of Seebeck coefficient can be achieved. The study has hence demonstrated that these vertical devices are very promising for enlarging the applications of graphene in transistors, strain sensors and thermoelectric devices.

\textbf{\textit{Acknowledgment.}} This research in Hanoi is funded by Vietnam's National Foundation for Science and Technology Development (NAFOSTED) under grant number 103.01-2014.24.

\end{document}